\documentclass{article}
\usepackage{spconf,amsmath,graphicx}
\usepackage[numbers,sort&compress]{natbib}

\usepackage{epsfig}
\usepackage{amsfonts}
\usepackage{algorithm}
\usepackage{algorithmic}
\usepackage{multirow}
\usepackage{url}

\usepackage{hyperref}

\title{Referee: Towards reference-free cross-speaker style transfer with low-quality data for expressive speech synthesis}

\name{Songxiang Liu, Shan Yang, Dan Su, Dong Yu}
\address{Tencent AI Lab}

\begin{document}
%
\maketitle
\begin{abstract}
Cross-speaker style transfer (CSST) in text-to-speech (TTS) synthesis aims at transferring a speaking style to the synthesised speech in a target speaker's voice.
Most previous CSST approaches rely on expensive high-quality data carrying desired speaking style during training and require a reference utterance to obtain speaking style descriptors as conditioning on the generation of a new sentence.
This work presents Referee, a robust reference-free CSST approach for expressive TTS, which fully leverages low-quality data to learn speaking styles from text.
Referee is built by cascading a text-to-style (T2S) model with a style-to-wave (S2W) model.
Phonetic PosteriorGram (PPG), phoneme-level pitch and energy contours are adopted as fine-grained speaking style descriptors, which are predicted from text using the T2S model. A novel pretrain-refinement method is adopted to learn a robust T2S model by only using readily accessible low-quality data.
The S2W model is trained with high-quality target data, which is adopted to effectively aggregate style descriptors and generate high-fidelity speech in the target speaker's voice.
Experimental results are presented, showing that Referee outperforms a global-style-token (GST)-based baseline approach in CSST.

\end{abstract}
\section{Introduction}
\label{sec1:intro}

Neural text-to-speech (TTS) synthesis has been greatly improved in terms of quality and robustness of synthesized speech in recent years \cite{wang2017tacotron,shen2018natural,yu2019durian,ren2019fastspeech}.
Nowadays, how to improve the expressiveness of TTS systems for an even better listening experience has attracted more attention and research effort.
Cross-speaker style transfer (CSST) is a promising technology for expressive speech synthesis, which aims at transferring a speaking style from a source speaker to the synthesized speech in a target speaker's voice.

Most previous CSST approaches require a reference utterance conveying the desired speaking style to obtained style descriptors, either in utterance level \cite{skerry2018towards,wang2018style,hsu2018hierarchical} or in more fine-grained levels \cite{lee2019robust,klimkov2019fine,Karlapati2020}, during the run-time inference phase. This hinders their practical use since a reference utterance is not universally applicable for different textual input and should be carefully selected for the desired speaking style.
Hence, reference-free CSST approach for expressive TTS is more applicable in real-world scenarios.

Reference-free CSST presents several challenges.
First, representing speaking style in a computable form is challenging, since a speaking style is the confluence of several factors including pronunciation patterns, intonation, intensity, etc. Due to the fact that all kinds of speech information are entangled within an acoustic signal, for robust CSST it is of great importance to decompose style representations from speaker identity.
Second, it is difficult to predict style descriptors from the text. Sufficient amount of training data conveying the desired speaking style is required to learn a robust text-to-style model.
Furthermore, it is challenging for a synthesis module to effectively and efficiently aggregate style descriptors and generate high-fidelity speech in a desired target speaker's voice.

Few work focuses on reference-free CSST approach.
In \cite{pan2021cross}, an auto-regressive multi-speaker and multi-style Transformer-based TTS model is augmented with a phoneme-level prosody module, and is able to conduct reference-free CSST. However, a sufficient amount of high-quality style data from non-target speakers are required during the training phase.

\begin{figure}[t]
	\centering
	\includegraphics[width=0.48\textwidth]{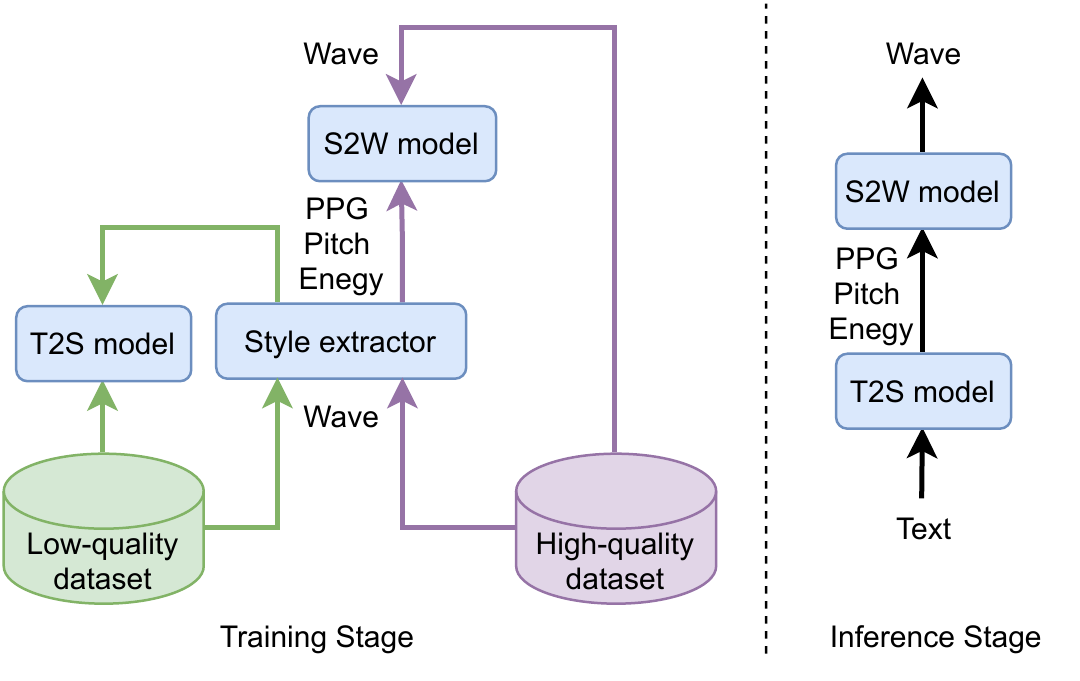}
	\caption{Training and inference procedures of the proposed CSST approach for TTS.}
	\label{fig:train_inference}
\end{figure}

This work attempts to address the aforementioned challenges, and proposes Referee, a robust \textbf{refer}ence-fr\textbf{ee} CSST approach.
An overview of the proposed approach is illustrated in Fig.~\ref{fig:train_inference}.
We regard pronunciation, duration, pitch, and intensity as basic information characterizing a speaking style.
Phonetic PosteriorGram (PPG) has been widely used to represent pronunciation in speech synthesis tasks due to its speaker-independent property, such as cross-lingual TTS \cite{sun2016personalized}, code-switched TTS \cite{cao2020code}, accent conversion \cite{zhao2018accent}, one-shot voice conversion \cite{liu2018voice}, just to name a few. A PPG is obtained from the acoustic model in a speaker-independent automatic speech recognition (SI-ASR) system, by stacking a sequence of phonetic posterior probabilistic vectors. Since PPG features have the same frame-rate as their corresponding acoustic features, they contain not only pronunciation information but also duration information of a speech signal. Moreover, PPGs computed from an SI-ASR system, which is trained on large-scale speech corpus containing various levels environmental noise, reverberation and any other kinds of acoustic distortions, are robust representations for pronunciation and duration information.
Therefore, this work uses PPGs as pronunciation and duration descriptors. 
Phoneme-level pitch and energy contours are utilized to represent pitch and intensity of a speech signal, respectively. In Referee, a style extractor is adopted to obtain these features from wave signals, as shown in Fig.~\ref{fig:train_inference}.

For reference-free CSST, this work trains a text-to-style (T2S) model to predict PPGs, phoneme-level pitch, and energy contours from the text. 
Large-scale high-quality dataset of a desired speaking style is expensive if not impossible to collect. 
In the meantime, we observe that a large number of low-quality speech data with rich speaking styles can be easily obtained in the era of data explosion. 
Based on this, we first pretrain a base T2S model with a multi-style low-quality dataset, and then refine it with the low-quality samples carrying the desired style we want to transfer with a novel meta-learning mechanism, as illustrated by the green arrows in Fig.~\ref{fig:train_inference}.
Finally, we train a style-to-wave (S2W) model with high-quality data from the target speaker to aggregate PPGs, pitch and energy (see purple arrows in Fig.~\ref{fig:train_inference}) and use it to generate speech carrying desired speaking style in the target speaker's voice.

The contributions of this work are summarized as follows:
\begin{itemize}
    \item We achieve robust reference-free CSST for expressive speech synthesis by cascading a T2S model with a powerful S2W model.
    \item We present a novel pretrain-refinement method to fully exploit readily accessible low-quality data.
\end{itemize}

The rest of this paper is organized as follows:
Details of the proposed approach are presented in Section~\ref{sec3}.  Experimental results are shown in Section~\ref{sec4} and Section~\ref{sec5} concludes this paper.


\section{Proposed approach}
\label{sec3}

The proposed approach consists of a style extractor, a T2S model and an S2W model, as illustrated in Fig.~\ref{fig:train_inference}. The style extractor is used to obtain PPGs, phoneme-level pitch and energy contours from wave signals. The T2S model is trained with low-quality data and adopted to predict PPGs, pitch and energy from text, while the S2W is trained with high-quality data from the target speaker and adopted to generate waveform in the target speaker's voice from PPG, pitch and energy.

\subsection{Style Extractor}

The style extractor contains a PPG model, a pitch extractor and an energy extractor. 
A Deep-FSMN with Self-Attention (DFSMN-SAN)-based ASR acoustic model \cite{you2020dfsmn} is trained as the PPG model with large-scale (about 20k hours) forced-aligned audio-text speech data, which contains diverse levels of pronunciation variations, environmental noises, reverberations, and other acoustic distortions. We train the ASR model with a frame-wise cross-entropy loss, where the ground-truth labels are Chinese phonemes (i.e., initial and final with tone).
The pitch extractor and energy extractor compute phoneme-level F0 values and energy values according to text-audio alignment information obtained from a hidden-Markov-model (HMM)-based forced aligner. Ground-truth frame-level F0 and energy  contours are averaged by durations to get phoneme-level pitch and energy.

\subsection{Text-to-Style Model}

\begin{figure}[t]
	\centering
	\includegraphics[width=0.40\textwidth]{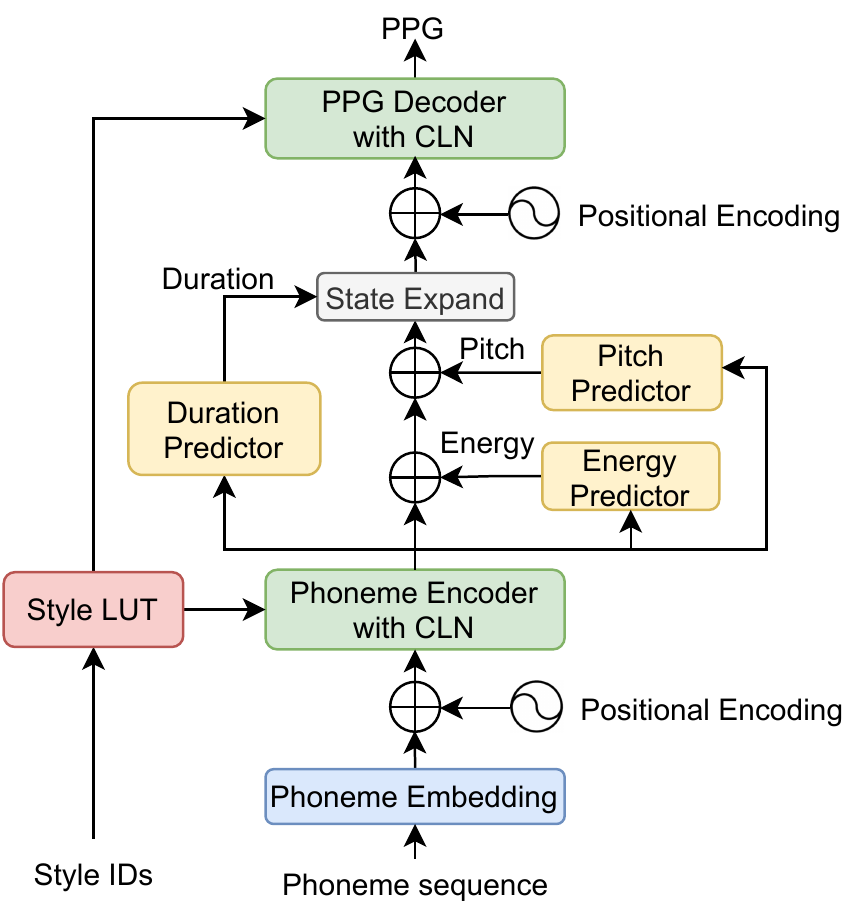}
	\caption{Text-to-Style (T2S) model. ``LUT'' is abbreviation of ``look-up table''. ``CLN'' is abbreviation of ``conditional layer normalization''.}
	\label{fig:t2s}
\end{figure}

The network architecture of the T2S model is illustrated in Fig.~\ref{fig:t2s}, which is adapted from FastSpeech 2 \cite{ren2021fastspeech} and further augmented with conditional layer normalization (CLN) introduced in AdaSpeech \cite{chen2021adaspeech}.
The T2S model takes phoneme sequence as input and predicts style descriptors (i.e., PPGs, phoneme-level pitch and energy contours) conditioned on style IDs.
Learning to predict style descriptors purely from textual input requires considerable amount of training data. To ease the inconvenience of collecting high-quality speech data, we fully exploit low-quality data, and design a novel pretrain-refinement scheme to learn a robust T2S model for a speaking style we want to transfer. The details will be presented in following subsections.

\subsubsection{T2S pretraining}

We use a low-quality multi-style audio-book corpus (each speaker has only one style) to train a T2S model as the base model. 
We incorporating style information into both the phoneme encoder and the PPG decoder via the operations of CLN, since we hypothesize that the phoneme encoder output contains not only content information but also pronunciation patterns.
Specifically, style IDs are first transformed to embedding vectors by a style look-up table (LUT). The style embedding vectors are then consumed by layer-dependent linear layers to generate scale and bias vectors, which are used to modulate the hidden features at the layer normalization layers after each self-attention and feed-forward network in the phoneme encoder and PPG decoder.
During pretraining, ground-truth phoneme-level pitch and energy features are added to the phoneme encoder output. Besides, ground-truth duration information is adopted to expand the encoder output along the temporal axis.
Mean squared error (MSE) criterion is adopted to compute PPG, pitch, energy and duration (in log-scale) losses.

\subsubsection{T2S refinement}
\label{t2s-refine}

After pretraining, we refine the base T2S model to a target style with a novel refinement procedure using a meta-learning mechanism.
One challenge of CSST is that when conducting style transfer for out-of-domain (OD) textual input, the transfer performance degrades to some degree compared with the cases where the textual input is in-domain. This is because text and style are somewhat entangled due to data sparsity in the training set.
To mitigate this issue, during the refinement stage, we make the T2S model simulate the scenario of conducting style transfer for OD texts using episodic meta-learning, using a similar but different procedure as shown in Meta-StyleSpeech \cite{min2021meta}.

\begin{figure}[t]
	\centering
	\includegraphics[width=0.48\textwidth]{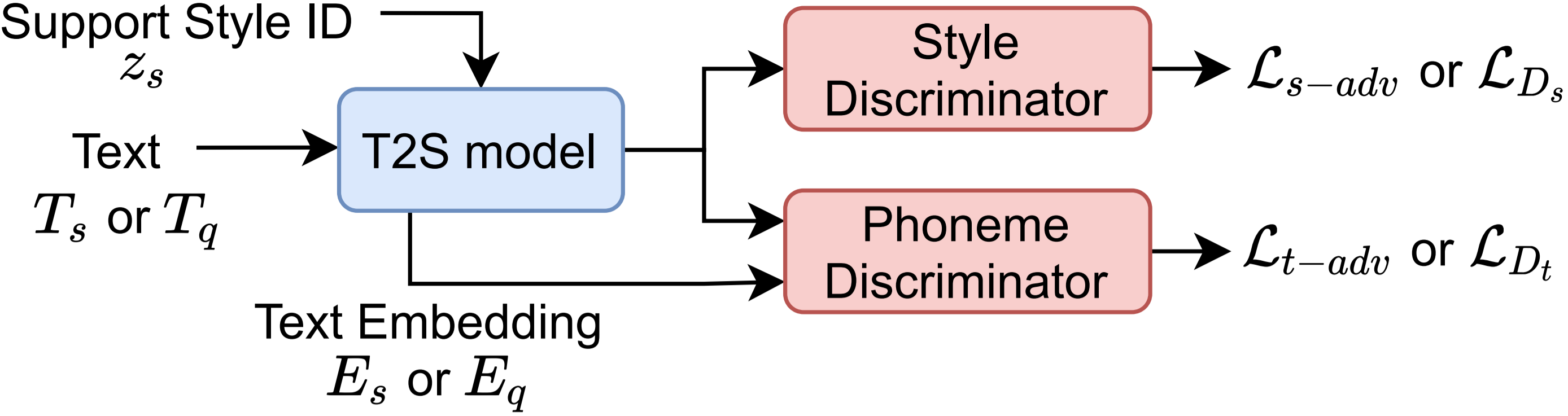}
	\caption{Refinement process of the S2T model. The reconstruction loss is not shown for simplicity.}
	\label{fig:refinement}
\end{figure}

Refinement process of the S2T model is illustrated in Fig.~\ref{fig:refinement}. For convenient presentation, we denote the tuple of paired ground-truth (PPG, pitch, energy) as $S$ and the predicted by the T2S model as $\hat{S}$.
We assume that only a single text sample for a speaking style is available; thus, simulating one-shot learning for new OD text input. In each episodic, we regard the target style as the support style ID $z_s$ and sample a support text $T_s$ in the target style.
In addition, we sample a query text $T_q$ from other styles (i.e., non-target styles).
Since there is no ground-truth $S$ for the query text in the target style, we set a style discriminator $D_s$ and a phoneme discriminator $D_t$ as shown in Fig.~\ref{fig:refinement} to give learning signals to the T2S model.
Due to the fact that the Transformer-base T2S model is parameter-intensive, we only update the parameters of the style embedding, CLN linear layers, pitch predictor, energy predictor and duration predictor; and freeze the remaining parameters in the T2S model during refinement. This helps mitigate the overfitting issue.
The discriminators $D_s$ and $D_t$ are set in the least squares GAN (LS-GAN) \cite{mao2017least} setting. Their loss function is computed as follows:
\begin{equation}
    \mathcal{L}_{D_s} = \mathbb{E}[(D_s(S) -1)^2 + D_s(T2S(T_q, z_s))]
\end{equation}
\begin{equation}
    \mathcal{L}_{D_t} = \mathbb{E}[(D_t(S, E_s) -1)^2 + D_t(T2S(T_q, z_s), E_q)]
\end{equation}
The adversarial loss of the T2S model is the sum of the adversarial losses from $D_s$ and $D_t$, as follows:
\begin{equation}
\begin{split}
&\mathcal{L}_{adv} = \mathcal{L}_{s-adv} + \mathcal{L}_{t-adv} \\ 
&= \mathbb{E}[(D_s(T2S(T_q, z_s)) - 1)^2] + \mathbb{E}[(D_t(T2S(T_q, z_s), E_q) - 1)^2]
\end{split}
\end{equation}
To stabilize the refinement process, we also use the reconstruction loss (e.g., MSE loss) for the T2S model as:
\begin{equation}
    \mathcal{L}_{recon} = \mathbb{E}[||T2S(T_s, z_s) - S||_2^2]
\end{equation}
Therefore, the ultimate losses of the T2S model and the discriminators are as follows:
\begin{equation}
    \mathcal{L}_{T2S} = \alpha \mathcal{L}_{recon} + \mathcal{L}_{adv}
\end{equation}
\begin{equation}
    \mathcal{L}_{D} = \mathcal{L}_{D_s} + \mathcal{L}_{D_t}
\end{equation}
We empirically set $\alpha=10$ to balance scale of each loss, and alternatively update the T2S model and the discriminators.

\begin{figure}[t]
	\centering
	\includegraphics[width=0.45\textwidth]{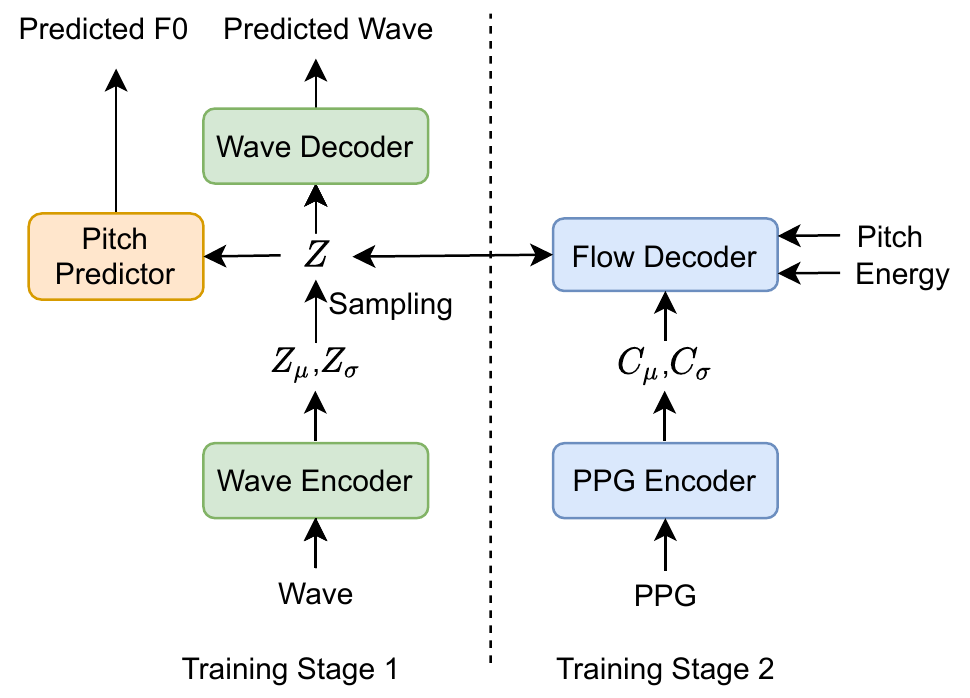}
	\caption{Training procedure of  the style-to-wave (S2W) model.}
	\label{fig:s2w}
\end{figure}

\subsection{Style-to-Wave Model}

The S2W model aggregates PPG, pitch, energy features and directly generate speech waveform in the target speaker's voice. We adapt the powerful Glow-WaveGAN model \cite{cong2021glow} as the S2W model, inspired by its success in high-fidelity TTS. We make several changes from the original structure to fit our need in this work: (1) The text encoder is replaced with a PPG encoder; (2) We drop the monotonic alignment search (MAS) module since PPGs are automatically aligned frame-wisely with the wave encoder output $Z_{\mu}$ and $Z_\sigma$, as well as the sampled hidden variable $Z$; and (3) we locally condition the Flow-based decoder with frame-wise pitch and energy features, obtained by expanding their phoneme-level counterparts according to phoneme durations. The training process of the S2W model consists of two stages, as illustrated in Fig.~\ref{fig:s2w}. We summarize them as follows, for more detailed description, we refer readers to \cite{cong2021glow}: At training stage 1, a generative adversarial network (GAN)-based variational autoencoder (VAE) is trained on the raw waveform, with an auxiliary task of predicting F0 from the latent variable $Z$;  At training stage 2, the wave encoder is frozen and used as a hidden variable sampler to train the PPG encoder and Flow decoder.

\subsection{Inference}

The inference procedure for CSST is illustrated in the right part of Fig.~\ref{fig:train_inference}. Text with arbitrary content is fed into the T2S model, which is refined on the target style, to predict PPG, phoneme-level pitch, energy and duration. The phoneme-level pitch and energy are then expanded to frame-level according to the predicted duration. Finally, the S2W model takes the predicted PPG and frame-level pitch and energy as input and generate waveform in the target speaker's voice conveying the target style; thus accomplishes reference-free CSST. 

\section{Experiments}
\label{sec4}

\subsection{Setups}
\label{data_prep}
Two internal Chinese corpora are used in the experiments. The low-quality multi-style audio-book corpus contains 38 speakers (each has only one speaking style) and in total has 28.8 hours speech data (0.76 hours/speaker). Sampling rate of some audio samples is 16 KHz. We resample all audio to 24 KHz. We randomly partition the low-quality corpus into training, validation and test sets according to an 86\%-7\%-7\% scheme. In the experiments, we choose a strong northeast Chinese accent as our target style, whose corresponding speaker is a male speaker.
The high-quality corpus is recorded for developing TTS systems by a female target speaker in reading style, which contains 14 hours of speech data. Audio is sampled at 24 KHz.
PPG feature has dimensionality of 218 with a 10ms frame-shift.

The T2S model in the proposed approach uses 4 feed-forward Transformer (FFT) blocks in both the phoneme encoder and the PPG decoder. The hidden size, number of attention heads, kernel size and filer size of the 1D convolution in the FFT block are set as 256, 2, 9 and 1024, respectively. The dimention of the style LUT is 128. The style discriminator and phoneme discriminator adopt the same network hyper-parameter settings as in \cite{min2021meta}. The S2W model follows the basic setting as that used in \cite{cong2021glow}. The frame-level pitch and energy features are incorporated into the Flow decoder by transforming with 1D convolutional layers with kernel size 3 and adding elementwisely to the hidden features in each affine coupling layer as local conditioning. The Adam optimizer with $\beta_1=0.9$, $\beta_1=0.98$ and $\epsilon=10^{-9}$ is used to train all the models. The T2S model is pretrained on the multi-style low-quality data with a batch size of 64 for 50K steps. Then we meta-refine the base T2S model on the target style for 5K steps using a fixed learning rate of $10^{-5}$ and batch size of 16.


\subsection{Comparisons}
Four approaches are compared in the experiments: \\
\textbf{GST-DurIAN} (baseline) This model is base on the DurIAN model \cite{yu2019durian}, which is a text-to-mel-spectrogram model augmented with a global-style-token (GST)-based style encoder \cite{wang2018style}. We use style vector from the GST encoder and speaker IDs as auxiliary input to both the acoustic model and duration model. To disentangle style and speaker, we adopt the domain adversarial training mechanism \cite{ganin2016domain}. GST-DurIAN is first pre-trained on the combination of both the low-quality multi-style corpus and the  high-quality target speaker corpus, and then refined to the low-quality target style data and the high-quality target speaker corpus. 
During cross-speaker style transfer at inference, a reference audio conveying the target style is used as the input to the GST encoder.
A HifiGAN vocoder \cite{hifigan} is used to generate waveform from mel-spectrogram. \\
\textbf{Style-GWG} (topline) This model (\textbf{Style}-\textbf{G}low-\textbf{W}ave\textbf{G}AN) is the topline of the proposed approach. We directly feed ground-truth PPG, pitch and energy obtained from a reference utterance to the S2W model. \\
\textbf{Prop-1} (proposed) This model is the proposed approach with a different refinement procedure. During the refinement, we use only the target style data to refine the T2S model. To avoid overfitting, only parameters of the style embedding, CLN linear layers, pitch predictor, energy predictor and duration predictor are updated.  \\
\textbf{Referee} (proposed) The proposed approach introduced in Section~\ref{sec3}. \\

\subsection{Evaluations}

We conduct the mean opinion score (MOS) tests and similarity MOS (SMOS) tests to evaluate the naturalness and voice similarity of the generated audio samples.
To evaluate the style transfer performance, ABX style perceptual preference tests are conducted between compared approaches. 10 samples are generated from hold-out texts each compared approach.
We invite 10 native Mandarin Chinese speakers as the raters, each presented with 370 stimuli on average.
Audio samples can be found online\footnote{[Online] \url{https://liusongxiang.github.io/Referee/}}.

\begin{table}[]
\caption{Mean opinion score (MOS) and similarity MOS (SMOS) results with 95\% confidence interval.}
\resizebox{0.48\textwidth}{!}{
\begin{tabular}{|c|c|c|c|c|}
\hline
\multirow{2}{*}{Setting}&\multicolumn{2}{c|}{In-domain text}&\multicolumn{2}{c|}{Out-of-domain text} \\ \cline{2-5} 
&\multicolumn{1}{c|}{MOS ($\uparrow$)}&\multicolumn{1}{c|}{SMOS($\uparrow$)}&MOS ($\uparrow$)&SMOS($\uparrow$)\\ \hline
Recordings  & 4.60$\pm$0.09 & \multicolumn{1}{c|}{-}                   & \multicolumn{1}{c|}{-} & \multicolumn{1}{c|}{-} \\ \hline
GST-DurIAN  & 3.38$\pm$0.14 & 3.53$\pm$0.19   & 3.27$\pm$0.21   & 3.25$\pm$0.18      \\ \hline
Style-GWG   & \bf{4.04$\pm$0.17} & \bf{3.67$\pm$0.22}   & -   & -      \\ \hline
Prop-1      & 3.85$\pm$0.17 & 3.55$\pm$0.20   & 3.33$\pm$0.20   & 3.23$\pm$0.19      \\ \hline
Referee     & 3.86$\pm$0.19 & 3.53$\pm$0.21   & \bf{3.35$\pm$0.21}   & \bf{3.33$\pm$0.22}      \\ \hline
\end{tabular}}
\label{tab:1}
\end{table}

\begin{figure}[t]
	\centering
	\includegraphics[width=0.48\textwidth]{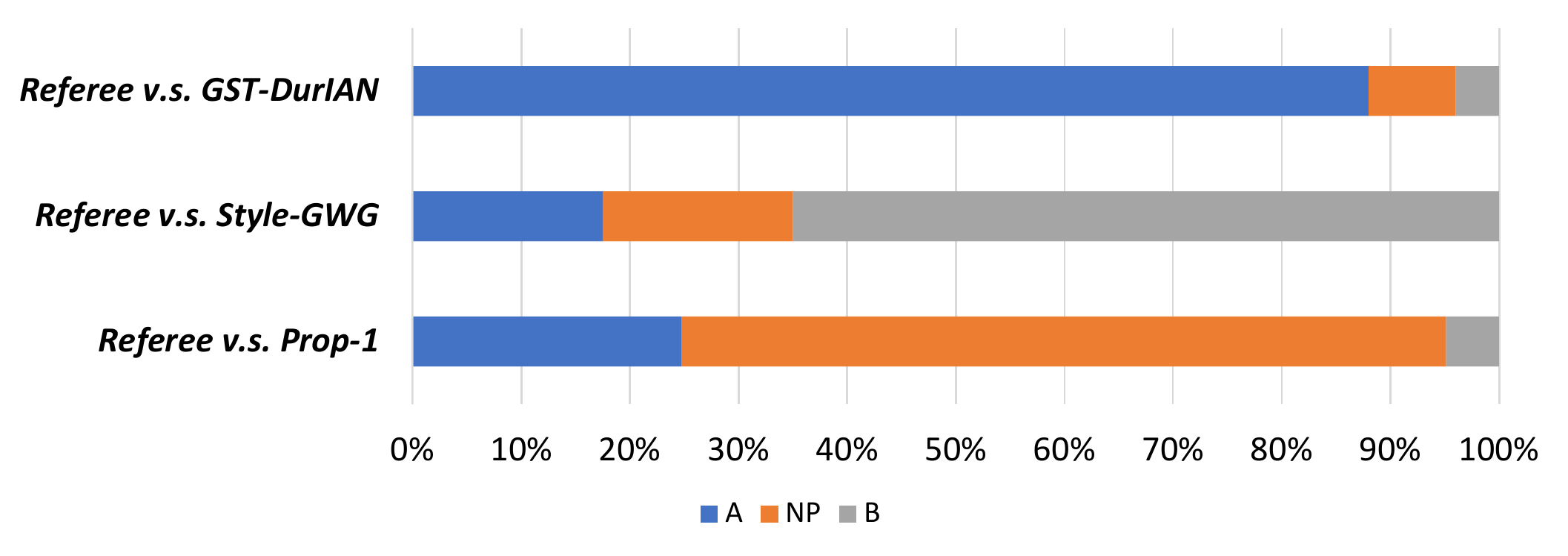}
	\caption{In-domain Style similarity ABX test results.}
	\label{fig:abx-id}
\end{figure}

\begin{figure}[t]
	\centering
	\includegraphics[width=0.48\textwidth]{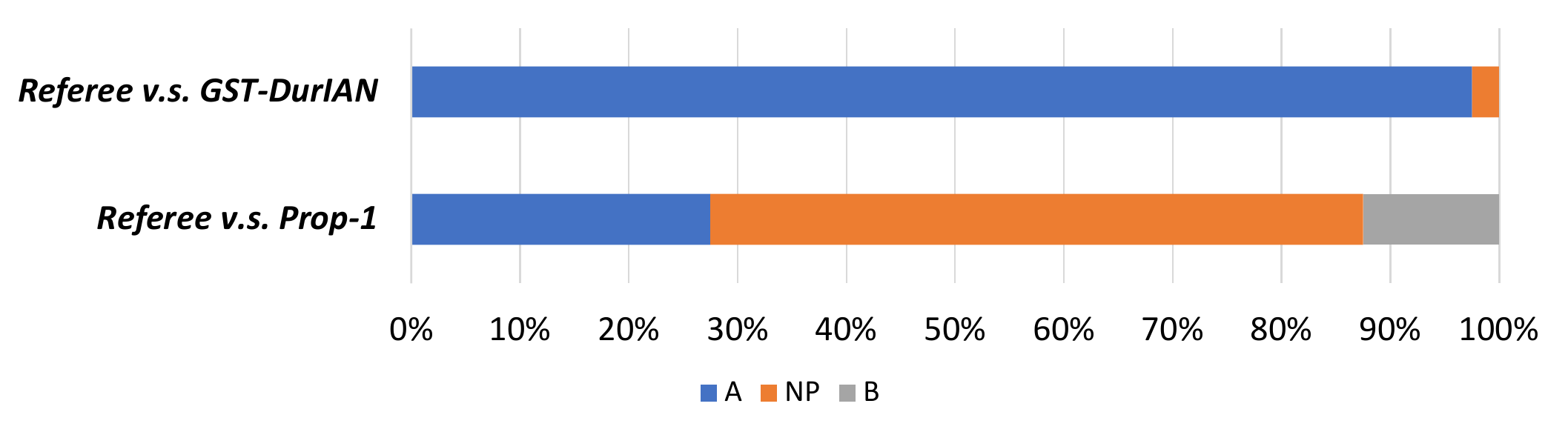}
	\caption{Out-domain Style similarity ABX test results.}
	\label{fig:abx-od}
\end{figure}

\subsubsection{CSST for in-domain text}

For the in-domain evaluation, texts from the test set in target style are used. The MOS and SMOS results are presented in the left part of Table~\ref{tab:1}.
We can see that the topline approach Style-GWG achieves the best MOS and SMOS results among the four compared approaches. Prop-1 and Referee have similar performance in naturalness and voice similarity of the generated samples, both outperforming the baseline approach GST-DurIAN in naturalness.
Since low-quality data is incorporated in the training process of the GST-DurIAN model, the style vector computed from a low-quality utterance by the GST encoder also encodes noise and channel information, leading to quality degradation in generated speech.
The ABX test results for evaluating style transfer performance are shown in Fig.~\ref{fig:abx-id}. We can see that Referee is significantly better than Prop-1 ($p$-value $\approx$ 0.016) and the baseline GST-DurIAN ($p$-value $= 2.0e^{-17}$) in terms of style transfer performance.
The topline Style-GWG beats Referee with a large margin ($p$-value = 0.00023).
Aggregating the observations mentioned above, we can safely say that Referee achieves significantly better performance than GST-DurIAN in terms of naturalness and cross-speaker style transfer; and on-par performance in terms of voice similarity.
Adopting the pretrain-refinement scheme introduced in section~\ref{t2s-refine} boosts Referee in naturalness and style transfer, compared with Prop-1.
However, there is still a gap between Referee and the topline Style-GWG. One possible reason is that there exists mismatch between the generated style features by the T2S model and those directly extracted from waveform. Finetuning the S2W model with generated style features should be the first attempt to address this issue.


\subsubsection{CSST for out-of-domain text}

This subsection presents the out-domain evaluation, where texts for generating samples are from an unseen reading-style corpus.
Since Style-GWG cannot conduct style transfer when there is no speech sample accessible for the target style, evaluations on it are not conducted.
The MOS and SMOS results are shown in the right part of Table~\ref{tab:1}. The results have a significant drop compared to their in-domain counterparts, showing that the models overfit the text in the training set in some degree. Enlarging the training set size could be a good solution to this issue.
Referee achieves the best results in naturalness and voice similarity.
The ABX test results are shown in Fig.~\ref{fig:abx-od}, where we see that Referee is significantly better than GST-DurIAN ($p$-value $= 3.7e^{-33}$). Adopting the pretrain-refinement scheme introduced in section~\ref{t2s-refine} helps Referee achieve slightly better style transfer performance compared with Prop-1 ($p$-value $= 0.068$).

\section{Conclusion}
\label{sec5}

This paper has introduced Referee, which is a robust CSST approach and does not require a reference stylish utterance at inference. PPG, phoneme-level pitch and energy are used as style descriptors and are predicted from text using a T2S model. Referee adopts a novel pretrain-refinement scheme to fully exploit easily accessible low-quality data.
Our experiments show that Referee outperforms a GST-based baseline system in CSST performance for both in-domain text and out-of-domain text. Future work includes extending Referee for supporting robust many-to-many cross-speaker style transfer.
\bibliographystyle{IEEEbib_abbrev}

\bibliography{strings,refs}

\end{document}